\documentclass[hyper]{JHEP} 
\usepackage{graphicx,amssymb,bm,latexsym,amsmath}

\title{Oscillating Strings and Non-Abelian T-dual Klebanov-Witten Background}

\author{Pabitra M. Pradhan\\
Department of Physics, Indian Institute of Technology Kharagpur,\\
Kharagpur-721 302, India \\
Email: \email{ppabitra@phy.iitkgp.ernet.in}} \vskip .2in
\abstract{We study oscillating string solutions in the
Klebanov-Witten and its non-Abelian T-dual background dualised
along an SU(2) isometry. We find the string energy as the function
of oscillation number and angular momentum. We show that for a
particular set of T-dual co-ordinates both the background have
equal string states. We also study the string states where the
strings are expanding and contracting in the T-dual co-ordinate
direction. We expect the presence of the superconformal field
theory dual operators whose anomalous dimensions depend on T-dual
co-ordinate.} 

\begin{document}

\section{Introduction}
Research on string dualities has added much of our understandings
to string theory. Establishing the duality between seemingly
theories, has been a major research area since the inception of
string theory. This has given many interesting and useful results,
from which few are mentioned below. First, it has taught us how to
relate various string theories in different regimes of validity
and compactifications. It also has led us to the discovery of
higher dimensional objects, such as D-branes and membranes.
Moreover, it has given us new insights to the non-perturbative
regime of the theory. The discovery of D-brane and its worldvolume
gauge theory, has prompted to propose an example of gauge/gravity
duality \cite{Maldacena:1997re}-\cite{Witten:1998qj}, which
relates the type IIB string theory in $AdS_5 \times S^5$
background to the $\mathcal N=4$ supersymmetric SU(N)Yang-Mills
gauge theory in four dimension. This duality maps the anomalous
dimensions of gauge-invariant operators in the gauge theory to the
energy spectrum of the string-theory states. This equivalence
beyond the Bogomol'nyi-Prasad-Sommerfield limit relies on the fact
that on the string-theory side the quantum corrections of strings
are suppressed by the large quantum number, while on the
field-theory side the anomalous dimension matrices of the dual
composite operators are related to the Hamiltonian of the
integrable spin chain. The idea of the operators with large
quantum number first proposed in \cite{Berenstein:2002jq} and
further explored in \cite{Gubser:2002tv}, which has prompted the
research on the semiclassical analysis of the rigidly rotating
string and its implications in the AdS/CFT correspondence. The
duality has been generalized to different models with
less-supersymmetric backgrounds with or without conformal
invariance \cite{Aharony:2002up}-\cite{Bigazzi:2003ui}. In the
semiclassical limit, both the rotating and oscillating strings
have been studied in both AdS and non-AdS , supersymmetric and
less-supersymmetric backgrounds
\cite{Engquist:2003rn}-\cite{Pradhan:2013sja}. Though the the
oscillating strings have more stability \cite{Khan:2005fc} than
the non-oscillating one, they are less explored compared to
rotating strings.

The T-duality is a symmetry transformation that relates different
string backgrounds with some isometries. The idea of generalizing
T-duality to include non-Abelian isometry groups has been worked
out in \cite{Ossa:1992vc}. When isometry groups are non-Abelian,
we reached at non-Abelian T-duality which is a proven technique to
construct supergravity duals of strongly coupled field theories
\cite{Sfetsos:2010uq}-\cite{Itsios:2013wd} and interestingly these
backgrounds retain supersymmetry \cite{Itsios:2013wd},
\cite{Barranco:2013fza}. Klebanov-Witten background is good
example of it where dualising along SU(2) isometries provide us a
type IIA supergravity $\sigma$-model background. In non-Abelian
transformation the isometry is partially destroyed which can be
recovered as non-local symmetry in the $\sigma$-model and the
corresponding $\sigma$-models are canonically equivalent
\cite{Curtright:1994be}. The semiclassical analysis of strings in
Klebanov-Witten background has been studies in
\cite{Tseytlin:2002ny}-\cite{Arnaudov:2010by}. Our study is
motivated by the recent paper \cite{Zacarias:2014wta}, where the
rotating string solutions have been worked out in Klebanov-Witten
and its non-Abelian T-dual background. It has been shown that both
backgrounds enjoy an equivalent subsector of states depending on
the values of the T-dual coordinates. Here we wish to study
oscillating strings in the Klebanov-Witten and its non-Abelian
T-dual background and compare the results in different regimes.

The rest of the paper is organized as follows. In
section-\ref{sec2}, we analyze the oscillating strings in the
Klebanov-Witten background when the oscillation is in AdS and in
$T^{1,1}$ separately. We find the energy and oscillation number
dispersion relation. In section-\ref{sec3}, we analyze different
classes of oscillating string configurations in non-Abelian T-dual
Klebanov-Witten background. We also study the case of the
oscillating string when the oscillation is in the T-dual
co-ordinate direction. In section-\ref{sec4}, we conclude with
some discussion.

\section{Oscillating string in $AdS_5 \times T^{1,1}$} \label{sec2}

We start with the Klebanov-Witten background which is the infrared
limit of the theory on N coincident D3 branes placed at the
conical singularity of $M_4 \times C$ \cite{Klebanov:1998hh}.
\begin{eqnarray}
ds^2 = &R^2& \Big(-\cosh^2\rho dt^2 + d\rho^2 + \sinh^2\rho
(d\chi^2 + \sin^2 \chi d\xi^2_1 + \cos^2\chi d\xi^2_2)\Big) \cr &&
\cr &+& R^2 \Big(\lambda^2_1(\sigma^2_{\hat{1}} +
\sigma^2_{\hat{2}}) + \lambda^2_2(\sigma^2_1 + \sigma^2_2) +
\lambda^2(\sigma_3 + \cos\theta_1 d\phi_1)^2 \Big) , \label{1}
\end{eqnarray}
where $R^2$ is the curvature radius of $T^{1,1}, ~~ \lambda^2_1 =
\lambda^2_2 = \frac{1}{6},~~\lambda^2=\frac{1}{9}$,
\begin{eqnarray}
&&\sigma_{\hat{1}} = \sin\theta_1 d\phi_1, \hskip 2cm
\sigma_{\hat{2}} = d\theta_1, \hskip 2cm \sigma_3 = d\psi +
\cos\theta_2 d\phi_2, \cr && \cr && \sigma_1 = \cos\psi
\sin\theta_2 d\phi_2 - \sin\psi d\theta_2, \hskip 4em   \sigma_2 =
\sin\psi \sin\theta_2 d\phi_2 + \cos\psi d\theta_2, \nonumber
\end{eqnarray} and the chosen co-ordinates range as
$\rho \in [0,\infty],~~~\chi,\xi_i \in [0,2\pi],~~~\theta_i \in
[0,\pi],~~~\phi_i \in [0,2\pi],~~~ \psi \in [0, 4\pi]$ and $i =
1,2$. Here $T^{1,1}$ is the homogenous space $\frac{SU(2) \times
SU(2)}{U(1)}$ with the diagonal embedding of the U(1) and Einstein
metric to be $R_{ij} = 4g_{ij}$. This is an $\mathcal N = 1$
superconformal field theory which is dual to the Type IIB theory
compactified on $AdS_5 \times T^{1,1}$.

\subsection{Oscillating in AdS} \label{sec2.1}
Here we wish to study a class of string solution which is
oscillating in the radial $\rho$ direction of the AdS and at the
same time rotating along the $\psi$ direction of the $T^{1,1}$
with an angular momentum. So we chose our ansatz as follows
\begin{eqnarray}
&&\chi = \xi_i = 0, ~~~~~~~~t = t(\tau) , ~~~~~~~~~~~~~\rho = \rho
(\tau),\cr && \cr &&\theta_2 = \phi_i = 0,~~~~~~~~ \psi =
\psi(\tau), ~~~~~~~~~\theta _1 = \theta = m\sigma. \label{2}
\end{eqnarray}
Now putting above anastz in the equation (\ref{1}), we get the
relevant background as
\begin{equation}
ds^2 = R^2 \Big[- \cosh \rho dt^2 + d\rho^2 + \frac{1}{6} d
\theta^2 + \frac{1}{9} d \psi^2 \Big]. \label {3}
\end{equation}
For the above background the Polyakov action is written as
\begin{equation}
S_p = \frac{R^2}{4\pi} \int d\sigma d\tau \left[-\cosh^2 \rho
\dot{t}^2 +\dot{\rho}^2 - \frac{1}{6} m^2 + \frac{1}{9}
\dot{\psi}^2 \right], \label{4}
\end{equation}
where the `dot' denotes the derivative with respect to $\tau$. We
can write the equations of motion for t and $\rho$
\begin{eqnarray}
\ddot{t} + 2 \tanh\rho \dot{\rho} \dot{t} &=& 0, \cr && \cr
\ddot{\rho} + \cosh\rho \sinh\rho \dot{t}^2 &=& 0. \label{5}
\end{eqnarray}
Now from Virasoro constraint we get
\begin{equation}
\dot{\rho}^2 = \cosh^2\rho \dot{t}^2 - \frac{1}{6} m^2 -
\frac{1}{9} \dot{\psi}^2. \label{6}
\end{equation}
From the Polyakov action we get the conserved charges
\begin{eqnarray}
E &=& R^2 \mathcal E = R^2 \cosh^2\rho \dot{t}, \cr && \cr J &=&
R^2 \mathcal J = \frac{R^2}{9} \dot{\psi}. \label{7}
\end{eqnarray}
Now the equation (\ref{6}) changes to
\begin{equation}
\dot{\rho}^2 = \frac{\mathcal E}{\cosh^2\rho} -\kappa^2, \label{8}
\end{equation}
where $\kappa^2 = \frac{1}{6} m^2 + 9 \mathcal J^2$. The
oscillation number for string is
\begin{equation}
N = R^2 \mathcal N = \frac{R^2}{2\pi} \oint d\rho \dot{\rho}.
\label{9}
\end{equation}
By putting $x = \sinh^2 \rho$ in the above equation, we get
\begin{eqnarray}
\mathcal N = \frac{1}{\pi} \int _{0}^{a} dx \frac{\sqrt{\mathcal
E^2 - \kappa^2(1+x^2)}}{1+x^2}, \label{10}
\end{eqnarray}
where $a= \sqrt{\frac{\mathcal E^2 - \kappa^2}{\kappa^2}}$ and
$\mathcal E$ is $\mathcal E^2 > \kappa^2 > 0$. In order to compute
the oscillation number we take the partial derivative of the above
equation with respect to m as prescribed in \cite{Park:2005kt}
\begin{eqnarray}
\frac{\partial \mathcal N}{\partial m} &=& ~~- \frac{m}{6 \pi}
\int _{0}^{a} \frac{dx}{\sqrt{\mathcal E^2 - \kappa^2(1+x^2)}} \cr
&& \cr &=& ~~- \frac{m}{12 \kappa} =~~ - \frac{m}{12
\sqrt{\frac{1}{6}m^2 + 9 \mathcal J^2}}. \label{11}
\end{eqnarray}
For large $\mathcal E$, integrating the above equation over m
\begin{equation}
\mathcal N = \mathcal N_0 - \frac{1}{2} \sqrt{\frac{1}{6}m^2 + 9
\mathcal J^2}, \label{12}
\end{equation}
where the integration constant $\mathcal N_0 (\mathcal E, \mathcal
J)$ can be computed from the integral (\ref{10}) for m = 0.
\begin{eqnarray}
\mathcal N_0 &=&~~ \frac{1}{\pi} \int _{0}^{b} dx
\frac{\sqrt{\mathcal E^2 - \mathcal J^2(1+x^2)}}{1+x^2}\cr && \cr
&=& ~~\frac{1}{2} (\mathcal E - 3\mathcal J), \label{13}
\end{eqnarray}
where $b = \sqrt{\frac{\mathcal E^2 - 9 \mathcal J^2}{9 \mathcal
J^2}}$. Putting the above value in the equation(\ref{12}), we get
the energy as the function of oscillation number and angular
momentum as
\begin{equation}
\mathcal E = 2\mathcal N + 3\mathcal J - \sqrt{\frac{m^2}{6}+ 9
\mathcal J^2}. \label{14}
\end{equation}
The last term in the above equation can be expanded according to
the angular momentum. When $m^2 < 54 \mathcal J^2$,
\begin{equation}
\mathcal E = 2\mathcal N - 3 \mathcal J \left[\frac{1}{108}
\frac{m^2}{\mathcal J^2} - \frac{1}{23328}\frac{m^4}{\mathcal J^4}
+ \mathcal O[\frac{m^6}{\mathcal J^6}]\right]. \label{15}
\end{equation}
When $m^2 > 54 \mathcal J^2$,
\begin{equation}
\mathcal E = 2\mathcal N + 3 \mathcal J - \frac{m}{\sqrt{6}}
\left[1+ 27 \frac{\mathcal J^2}{m^2} - \frac{729}{2}
\frac{\mathcal J^4} {m^4} + \mathcal O[\frac{\mathcal J^6}
{m^6}]\right]. \label{16}
\end{equation}

\subsection{Oscillating in $T^{1,1}$}
In this subsection we study another class of string solution where
the string is oscillating in the $\theta$ direction of the
$T^{1,1}$. We chose the ansatz as
\begin{eqnarray}
&&t = t(\tau),~~~~~~ \rho = 0,~~~~~~ \theta_2 = 0,~~~~~~ \phi_2 =
0, \cr && \cr &&\theta_1 = \theta = \theta(\tau),~~~~~~ \phi_1 =
\phi = m\sigma,~~~~~~ \psi = 0. \label{17}
\end{eqnarray}
Now the metric in the equation (\ref{1}) changes to
\begin{equation}
ds^2 = R^2\Big[-dt^2 + \frac{1}{6} d\theta^2 + \left(\frac{1}{9} +
\frac{1}{18} \sin^2\theta\right) d\phi^2 \Big]. \label{18}
\end{equation}
We write the Polyakov action for the fundamental string in this
background
\begin{equation}
S_p = \frac{R^2}{4\pi} \int d\sigma d\tau \left[-\dot{t}^2 +
\frac{1}{6} \dot\theta^2 - \left(\frac{1}{9} + \frac{1}{18}
\sin^2\theta\right) m^2 \right]\label{19}
\end{equation}
The equation of motion for $\theta$ is
\begin{equation}
\ddot{\theta} + \frac{1}{3}m^2 \sin\theta \cos\theta = 0.
\label{20}
\end{equation}
But from Virasoro constraint we get
\begin{equation}
\dot{\theta}^2 = 6 \mathcal E^2 - \left(\frac{2}{3} + \frac{1}{3}
\sin^2\theta \right) m^2, \label{21}
\end{equation}
where $ E = R^2 \mathcal E$, is the energy. Now oscillation number
is
\begin{eqnarray}
\mathcal N &=& \frac{1}{2\pi} \oint d\theta \dot{\theta} \cr &&
\cr &=& \frac{1}{2\pi} \oint d\theta \sqrt{6 \mathcal E^2 -
(\frac{2}{3} + \frac{1}{3} \sin^2\theta) m^2}. \label{22}
\end{eqnarray}
Putting $\sin\theta = x$ in the above equation, we get
\begin{equation}
\mathcal N = \frac{1}{2\pi} \oint \frac{dx}{1-x^2} \sqrt{6
\mathcal E^2(1-x^2) - (\frac{2}{3} + \frac{1}{3} x^2) m^2(1-x^2)}.
\label{23}
\end{equation}
Differentiating the above equation with respect to m,
\begin{equation}
\frac{\partial \mathcal N}{\partial m} = - \frac{m}{2\pi} \oint
\frac{\frac{2}{3} + \frac{1}{3} x^2} {\sqrt{6 \mathcal E^2(1-x^2)
- (\frac{2}{3} + \frac{1}{3} x^2) m^2(1-x^2)}}dx. \label{24}
\end{equation}
We put $x^2 = y$ to compute the integral
\begin{equation}
\frac{\partial \mathcal N}{\partial m} = - \frac{m}{2\pi}
\int_{c}^{b} \frac{\frac{2}{3} + \frac{1}{3}y}{\sqrt{6\mathcal E^2
y (1-y) - (\frac{2}{3} + \frac{1}{3}y)m^2y(1-y)}} dy, \label{25}
\end{equation}
where $a, b, c$ are the roots of the polynomial of the denominator
of the above integral (\ref{25}). For large $\mathcal E$, we get a
lower bound to energy as $\mathcal E$ as $6\mathcal E^2 > m^2$ and
chose $a = \frac{18\mathcal E^2-2m^2}{m^2},~ b=1, ~c=0$.
\begin{equation}
\frac{\partial \mathcal N}{\partial m} = I_1 +I_2, \label{26}
\end{equation}
where
\begin{eqnarray}
I_1 &=& - \frac{m}{3\pi} \int_{c}^{b} \frac{dy}
{\sqrt{\frac{1}{3}(y-a)(y-b)(y-c)}} \cr && \cr &=& -\frac{m}{\pi}
\frac{2}{\sqrt{3a}} \mathbb{K}(1/a), \cr && \cr I_2 &=& -
\frac{m}{6\pi} \int_{c}^{b} \frac{y}
{\sqrt{\frac{1}{3}(y-a)(y-b)(y-c)}}dy \cr && \cr &=& \frac{m}{\pi}
\sqrt{\frac{a}{3}}~[\mathbb{E}(1/a)-\mathbb{K}(1/a)], \label{27}
\end{eqnarray}
where $\mathbb{K}$ and $\mathbb{E}$ are usual complete elliptic
integral of first and second kind respectively. We can expand
these elliptic integrals to get the energy as the function of
oscillation number.
\begin{equation}
\frac{\partial \mathcal N}{\partial m} = \frac{m}{\pi}
\sqrt{\frac{a}{3}}~\left[\mathbb{E}(1/a)-(1+\frac{2}{a})\mathbb{K}(1/a)\right].
\label{28}
\end{equation}
\begin{equation}
\frac{\partial \mathcal N}{\partial m} = -\frac{5}{12\sqrt{6}}
\frac{m^2}{\mathcal E} - \frac{17}{576\sqrt{6}}
\frac{m^4}{\mathcal E^3} - \frac{265}{82944\sqrt{6}}
\frac{m^6}{\mathcal E^5} + \mathcal O [\mathcal E^{-7}].
\label{29}
\end{equation}
Now, integrating the above equation over m from 0 to $\infty$ and
setting $\mathcal N (\infty) = 0$, we get
\begin{equation}
\mathcal N = \frac{5}{36\sqrt{6}} \frac{m^3}{\mathcal E} +
\frac{17}{2880\sqrt{6}} \frac{m^5}{\mathcal E^3} -
\frac{265}{580608\sqrt{6}} \frac{m^7}{\mathcal E^5} + \mathcal O
[\mathcal E^{-7}]. \label{30}
\end{equation}
Inversing the series we get the energy as
\begin{equation}
\mathcal E = 0.0567 m^3 \mathcal N^{-1} + 0.7495 m^{-1} \mathcal N
-1.789 m^{-5} \mathcal N^3 + 20.67 m^{-9} \mathcal N^5 + \mathcal
O [\mathcal N^7]. \label {31}
\end{equation}

\section{Oscillating string in Non-Abelian T-dual Klebanov- Witten
Background} \label{sec3} Here we take the dualised metric which is
presented in \cite{Itsios:2013wd} where dualisation is made with
respect to the SU(2) global isometry defined by the $\sigma_is$.
As $AdS_5 \times T^{1,1}$ is a block diagonal spacetime, $AdS_5$
comes as a spectator space. Also  two fields $\theta_1$ and
$\phi_1$ of $T^{1,1}$ come as spectator as the gauge choices are
taken accordingly. As the supersymmetry in the dual filed theory
of $AdS_5 \times T^{1,1}$ is uncharged under SU(2) flavour
symmetries,it is supposed to be persevered after the T-dualisation
along this SU(2). The resulted dualised background is
$\sigma-$model on a target space with $\mathcal N = 1 $
supersymmetric solution of type IIA.
\begin{eqnarray}
ds^2_{dual} = &&ds^{2}_{AdS_5} + \lambda^2_1(\sigma^2_{\hat{1}} +
\sigma^2_{\hat{2}}) + \frac{\lambda^2_2 \lambda^2}{\Delta} x^2_1
\sigma^2_{\hat{3}} \cr && \cr &&+ \frac{1}{\Delta}\Big((x^2_1 +
\lambda^2_2 \lambda^2) dx^2_1 +(x^2_2 + \lambda^4_2)dx^2_2 + 2x_1
x_2 dx_1 dx_2\Big), \label{32}
\end{eqnarray}
where $\sigma_{\hat{3}} = d\psi + \cos\theta_1 d\phi_1, ~~\Delta
\equiv \lambda^2_2 x^2_1 + \lambda^2 (x^2_2 + \lambda^4_2)$ and R
is taken to be 1 for convenience otherwise can be restored by the
suitable rescaling. For the small values of $x_1$ and fixed $x_2$
the metric on the internal space behaves as
\begin{equation}
ds^2_{dual} = ds^{2}_{AdS_5}+ \lambda^2_1(\sigma^2_{\hat{1}} +
\sigma^2_{\hat{2}}) + \frac{\lambda^2_2}{x^2_2 + \lambda^4_2}
x^2_1 \sigma^2_{\hat{3}}. \label{33}
\end{equation}
Though geometry is regular, above metric has a bolt singularity
which can be removed by changing the range of $\psi$ to be $2\pi$.

\subsection{Oscillating in AdS} \label{sec3.1}
Here, we study the solution for a string moving in the background
(\ref{33}) which is  oscillating in $AdS$ along $\rho$ and
simultaneously rotating along $\psi$ and localized at a fixed
point in the plane $(x_1, x_2)$. Our ansatz for this configuration
\begin{eqnarray}
&&t = t(\tau),~~~~~~~~~ \rho = \rho(\tau),~~~~~~~~~ \chi= \xi_i =
\phi_1 =0, \cr && \cr &&\theta_1 = m\sigma,~~~~~~~~~ \psi =
\psi(\tau),~~~~~~~~~ x_1,x_2 = fixed. \label{34}
\end{eqnarray}
Now the equation (\ref{33}) takes the form
\begin{eqnarray}
ds^2_{dual}= -\cosh^2\rho dt^2 +d\rho^2 + \lambda^2_1 d \theta^2_1
+ \frac{\lambda^2_2}{x^2_2 + \lambda^4_2} x^2_1 d\psi^2.
\label{35}
\end{eqnarray}
We write the Polyakov action for the above background (\ref{35})
\begin{equation}
S_p = \frac{1}{4\pi} \int d\sigma d\tau \Big(-\cosh^2\rho
\dot{t}^2 + \dot{\rho}^2 -m^2 \lambda^2_1 +
\frac{\lambda^2_2}{x^2_2 + \lambda^4_2} x^2_1 \dot{\psi}^2 \Big).
\label{36}
\end{equation}
From the Virasoro constraint we get
\begin{equation}
\dot{\rho}^2 = \frac{E^2} {\cosh^2\rho} -m^2 \lambda^2_1 -
\frac{x^2_2 + \lambda^4_2}{\lambda^2_2 x^2_1} J^2, \label{37}
\end{equation}
where E and J are conserved charges and can be computed as in the
section (\ref{sec2.1}). We can see the equations (\ref{8}) and
(\ref{37}) are equivalent. So the string states in (\ref{2}) and
(\ref{34}) are charatersied by the same labels (\ref{9}) to
(\ref{16}), if we choose appropriate values for $x_1$ and $x_2$.
This is possible in the finite range of $\frac{1}{3\sqrt{6}}< x_1
<1$ for suitable $x_2$ values which ranges as $0 < x_2 <
\frac{\sqrt{53}}{6}$.

\subsection{Oscillating in $\hat{T}^{1,1}$}
In this subsection, we study the string which moves in the
non-Abelian T-dual Klebanov-Witten background which is oscillating
along $\theta_1$ direction in $\hat{T}^{1,1}$ and localized at a
fixed point in the plane $(x_1, x_2)$. Our ansatz
\begin{eqnarray}
&&t = t(\tau),~~~~~~~~~ \rho = 0,~~~~~~~~~ \theta_1 = \theta =
\theta(\tau), \cr && \cr  &&\phi_1 = m\sigma,~~~~~~~~~ \psi =
0,~~~~~~~~~ x_1,x_2 = fixed. \label{38}
\end{eqnarray}
The relevant background looks like
\begin{equation}
ds^2_{dual} = -dt^2 + \lambda^2_1 d\theta^2 + \left[\lambda^2_1
\sin^2\theta + \frac{\lambda^2_2} {x^2_2+\lambda^4_2} x^2_1
\cos^2\theta\right] d\phi^2_1. \label{39}
\end{equation}
From Virasoro constraint, we get
\begin{equation}
\dot{\theta}^2 = \frac{E^2}{\lambda^2_1} - \left[\left(1-
\frac{\lambda^2_2} {\lambda^2_1(x^2_2+\lambda^4_2)}x^2_1\right)
\sin^2\theta + \frac{\lambda^2_2}
{\lambda^2_1(x^2_2+\lambda^4_2)}x^2_1 \right]m^2. \label{40}
\end{equation}
We can see that the equation (\ref{21}) is equivalent to
(\ref{40}) and labels (\ref{22}) to (\ref{31}) are same for the
string states described by (\ref{17}) and (\ref{38}) for the same
values of $x_1$ and $x_2$ as we got in the above subsection
(\ref{sec3.1}).

Now, we wish to study another class of oscillating string solution
where the string is fixed at some point $x_2$ and oscillating in
the T-dual co-ordinate direction $x_1$ from a minimum ($x_{1min}$)
to a maximum ($x_{1max}$) value. So, for this configuration our
ansatz
\begin{eqnarray}
&&t = t(\tau),~~~~~~~ \rho = 0,~~~~~~~ \theta_1 =0,~~~~~~~ \phi_1
= m\sigma, \cr && \cr  && \psi = 0,~~~~~~~~~~~
x_1=x_1(\tau),~~~~~~~~~~~x_2 = fixed. \label{41}
\end{eqnarray}
As $x_1$ is no more fixed, we use the background in the equation
(\ref{32}) and we get
\begin{equation}
ds^2_{dual} = -dt^2 + \frac{\lambda^2 \lambda^2_2}{\Delta} x^2_1
d\phi^2_1 + \frac{1}{\Delta} (x^2_1 + \lambda^2 \lambda^2_2)
dx^2_1. \label{42}
\end{equation}
From the Virasoro constraint we get
\begin{equation}
\dot{x_1}^2 = \frac{\Delta E^2 - \lambda^2 \lambda^2_2 m^2 x^2_1}
{x^2_1 + \lambda^2 \lambda^2_2}. \label{43}
\end{equation}
Now we write the oscillation number
\begin{equation}
N = \frac{1}{2\pi} \oint dx_1 \sqrt{\frac{\Delta E^2 - \lambda^2
\lambda^2_2 m^2 x^2_1} {x^2_1 + \lambda^2 \lambda^2_2}}.
\label{44}
\end{equation}
Using similar process to the previous section, we take the
derivative with respect to $m$ and put $x^2_1 = y$ to get
\begin{equation}
\frac{\partial N}{\partial m} = -\frac{m}{2\pi} \int_{r_2}^{r_3}
dy \frac{\lambda^2 \lambda^2_2 y}{\sqrt{y(y+\lambda^2
\lambda^2_2)(\Delta E^2 - m^2 \lambda^2 \lambda^2_2 y)}},
\label{45}
\end{equation}
Where $r_2,r_3$ are two positive roots of the polynomial in the
denominator of the above integral (\ref{45}). $r_1$ is the other
root which is a negative one. The above integral can be written in
terms of the usual complete elliptic integrals and further
expanded where the roots are $-\frac{1}{54}, 0$ and
$\frac{E^2(1+36 x^2_2)}{6m^2-54E^2}$.
\begin{eqnarray}
\frac{\partial N}{\partial m} &=& \frac{m}{\pi
\sqrt{54(m^2-9E^2)(r_3-r_1)}} \left[(r_1-r_3)\mathbb
{E}\left(\frac{r_3-r_2} {r_3-r_1} \right)-r_1 \mathbb {K}
\left(\frac{r_3-r_2} {r_3-r_1} \right) \right] \cr && \cr &=&
-0.04167(1+36 x^2_2) \frac{E^2}{m^2} + 0.42187 (-1-24 x^2_2
+432x^4_2) \frac{E^4}{m^4} \cr && \cr && - 0.79101 (5+108 x^2_2
-1296 x^4_2 +46656 x^6_2) \frac{E^6}{m^6} + \mathcal O [ E^8],
\label{46}
\end{eqnarray}
where $m^2>9E^2$ which this gives the upper bound to the energy
and in the $E^8$ term $x_2$ runs upto $x^8_2$ and so on. Now this
series can be integrated over m and inverted to get the energy
\begin{eqnarray}
E &=& \sqrt{\frac{\tilde{m}N}{(0.04167 + 1.5 x^2_2)^5}} \Big[1+
(-0.07031-1.6875 x^2_2 + 30.375 x^4_2)\frac{N}{\tilde{m}} \cr &&
\cr &&+ (0.00082-0.11865x^2_2 -13.526 x^4_2 -358.80 x^6_2 -2306.6
x^8_2) \frac{N^2}{\tilde{m}^2} + \mathcal O
\Big(\frac{N^3}{\tilde{m}^3}\Big)\Big], \label{47}
\end{eqnarray}
Where $\tilde{m} = m (0.04167 + 1.5 x^2_2)^2$. This is the energy
expression for the short strings which are oscillating in the
T-dual co-ordinate direction.

\section{Conclusion} \label{sec4}
In this paper, we have studied various oscillating strings in the
Klebanov-Witten and it Non-Abelian T-dual background background.
First we have studied the Oscillating long strings in $AdS_2
\times S^2$ where the string is expanding and contracting in the
radial direction of AdS and simultaneously has an angular momentum
in the $T^{1,1}$. We have found the energy as a linear function of
oscillation number. Then we have studied another class of
oscillating strings which are oscillating in $\mathbb {R} \times
T^{1,1}$ along the $T^{1,1}$. The energy for the long string comes
to be a series as a function of string oscillation number.

In the last section, we have studied the oscillating strings in
the non-Abelian T-dual Klebanov-Witten background. We fixed the
T-dual coordinates so that the remaining directions give a
squashed three sphere geometry and a reduction in the range of the
$\psi$ direction in order to remove the bolt singularity. In this
geometry, the T-dual co-ordinate $x_1=0$ singularity at fixed
$\theta_i, \phi_i$ is a co-ordinate singularity of $\mathbb {R}^2$
in polar co-ordinates. Here we have found both the Klebanov-Witten
and its T-dual background give the same result for a range of
T-dual co-ordinates though their field theory duals are different
in principle. Because a particular sector of geometry of $T^{1,1}$
is unaffected by the non-Abelian T-duality. As we see this is not
true through out the space as it is restricted by the range of
T-dual co-ordinates. Further we have studied some oscillating
strings which are oscillating in the T-dual co-ordinate direction
and found its energy expression for short string configuration. We
have also marked that though the R-charge ($\psi$)
\cite{Gaillard:2013vsa} vanishes there is non-vanishing T-dual
co-ordinate $x_1$ and solution in contrary to
\cite{Zacarias:2014wta}, where vanishing R-charge implies the
vanishing of T-dual co-ordinate $x_1$ for the rotating strings.
The solutions presented here are in the $AdS_5 \times T^{1,1}$ and
its T-dual background which are dual to the $\mathcal N = 1$
superconformal field theory with SU(2)$\times$ SU(2) flavor
symmetry. The chiral operators analogue of the operators in
$\mathcal N = 4$ SYM are given by Tr$(AB)^k$ with R-charge k and
in the $(\frac{k}{2},\frac{k}{2})$ representation of the flavor
group SU(2) $\times$ SU(2). Here the two chiral multiplets A and
B, which are elementary degrees of freedom, are correspondingly in
the (N, $\bar{N}$) and ($\bar{N}$ ,N) representations. We can
notice that the solution presented in the equation (\ref{47}) is
dependent upon the T-dual co-ordinate $x_2$ which prompts us to
expect the existence of superconformal field theory dual operator
to this string state whose anomalous dimension depend upon the
T-dual co-ordinate. However prediction of the exact form of the
operator for the solution in the equation (\ref{47}) can not be
done on basis of the work presented here. We leave this problem
for future work.

\section*{Acknowledgement}
PMP would like to thank S. Zacar\'{i}as for useful discussions and
K. L. Panigrahi for some useful comments.

\end{document}